\documentclass[pre,onecolumn,superscriptaddress,nofootinbib,showpacs,showkeys]{revtex4}

\usepackage{graphicx}
\usepackage{dcolumn}

\usepackage{eucal}
\usepackage[dvips]{epsfig}
\usepackage{amssymb}

\usepackage{amsmath,amsthm}

\usepackage[]{color}
\allowdisplaybreaks[4]

\newcommand{\Tr}{{\mathcal{T}r}}

\newcommand{\kB}{k_{\mathrm{B}}}

\newcommand{\SG}{S_{\mathrm{G}}}
\newcommand{\SB}{S_{\mathrm{B}}}
\newcommand{\SC}{S_{\mathrm{C}}}

\newcommand{\TG}{T_{\mathrm{G}}}
\newcommand{\TB}{T_{\mathrm{B}}}

\newcommand{\ZC}{Z_{\mathrm{C}}}

\newcommand{\ES}{E^{\mathcal{S}}}
\newcommand{\EB}{E^{\mathcal{B}}}
\newcommand{\ET}{E^{\mathcal{T}}}

\newcommand{\HS}{H^{\mathcal{S}}}
\newcommand{\HB}{H^{\mathcal{B}}}
\newcommand{\HT}{H^{\mathcal{T}}}

\newcommand{\SBB}{S^{\mathcal{B}}_{\mathrm{B}}}
\newcommand{\TBB}{T^{\mathcal{B}}_{\mathrm{B}}}
\newcommand{\CBB}{C^{\mathcal{B}}_{\mathrm{B}}}

\newcommand{\TGS}{T^{\mathcal{S}}_{\mathrm{G}}}
\newcommand{\TGB}{T^{\mathcal{B}}_{\mathrm{G}}}
\newcommand{\TGT}{T^{\mathcal{T}}_{\mathrm{G}}}

\newcommand{\rhoT}{\rho^{\mathcal{T}}}

\newcommand{\omegaB}{\omega^{\mathcal{B}}}
\newcommand{\omegaT}{\omega^{\mathcal{T}}}

\newcommand{\gS}{\omega^{\mathcal{S}}}

\newcommand{\EBE}{\bar{E}^{\mathrm{B}}}

\usepackage{color}

\begin{document}
%%%%%%%%%%%%%%%%%%%%%%%%%%%%%%%%%%%%%%%%%%%%%%%%%%%

\title{Meaning of temperature in different thermostatistical ensembles}
%\title{Entropy, Temperature, Thermal Equilibrium, and the Laws of Thermodynamics}

\author{Peter H{\"a}nggi}
\affiliation{Institute of Physics, University of Augsburg, Universit{\"a}tsstra{\ss}e 1, D-86135 Augsburg, Germany}
\affiliation{Nanosystems Initiative Munich, Schellingstr. 4, D-80799 M\"{u}nchen, Germany}

\author{Stefan Hilbert}
%\email{hilbert@mpa-garching.mpg.de}
\email{stefan.hilbert@tum.de}
\affiliation{Exzellenzcluster Universe, Boltzmannstr. 2, D-85748 Garching, Germany}
%\affiliation{Max-Planck-Institut f{\"u}r Astrophysik, Karl-Schwarzschild-Str. 1, D-85748 Garching, Germany}

\author{J{\"o}rn Dunkel}
\affiliation{Department of Mathematics, Massachusetts Institute of Technology, 77 Massachusetts Avenue E17-412, Cambridge, MA 02139-4307, USA}

\date{\today}

\begin{abstract}
Depending on the exact experimental conditions, the thermodynamic properties of physical systems can be related to one or more thermostatistical ensembles. Here, we survey the notion of thermodynamic temperature in different statistical ensembles, focusing in particular on subtleties that arise when ensembles become non-equivalent. The \lq mother\rq{} of all ensembles, the microcanonical ensemble, uses entropy and internal energy (the most fundamental, dynamically conserved quantity)  to derive temperature as a secondary thermodynamic variable.
Over the past century, some confusion has been caused by the fact that several competing  microcanonical entropy definitions are used in the literature, most commonly the volume  and  surface entropies introduced by Gibbs.  It can be proved, however, that only the volume entropy  satisfies exactly the traditional form of the  laws of thermodynamics for a broad class of physical systems, including all standard classical Hamiltonian systems, regardless of their size.  This mathematically rigorous fact implies that negative \lq absolute\rq{} temperatures and Carnot efficiencies $>1$ are not achievable within a standard thermodynamical framework. As an important offspring of microcanonical thermostatistics, we shall briefly consider the canonical ensemble and comment on the validity of the Boltzmann weight factor. We conclude by addressing open mathematical problems that arise for systems with discrete energy spectrum.
\end{abstract}



\maketitle

% pacs / keywords:
% 05.20.-y 	Classical statistical mechanics
% 05.20.Gg 	Classical ensemble theory

% 05.30.-d  Quantum statistical mechanics
% 05.30.Ch  Quantum ensemble theory

% 05.70.-a  Thermodynamics

\section{Introduction}

The fundamental differential relation~\cite{Kubo}
\begin{equation}
\label{e:T_def}
\frac{1}{T}=\frac{\partial S}{\partial E}
\end{equation}
connects the thermodynamic state functions temperature $T$, internal energy $E$, and entropy~$S$.
Given $S$ as a function of $E$, and possibly other control parameters, Eq.~\eqref{e:T_def} in fact  \emph{defines} the thermodynamic temperature.  The concept of entropy was introduced by Rudolf Clausius~\cite{Clausius} in 1865. Clausius chose the symbol $S$ in honor of (Nicolas L\'eonard) Sadi Carnot, who laid the groundwork for the Second Law of Thermodynamics. The celebrated Clausius relation $dS = \delta Q/T$ identifies the inverse of the thermodynamic temperature $T$ as the integrating factor for the Second Law, with $\delta Q\gtreqqless 0$  denoting quasi-static and reversible infinitesimal heat exchange. After Clausius' seminal paper~\cite{Clausius},  it took about 30 more years until Gibbs~\cite{Gibbs}, Einstein, Planck~\cite{Planck1,Planck2} and others~\cite{Hertz} were able to connect firmly thermodynamics  and statistical mechanics -- and yet certain aspects of this connection have remained a subject of debate up to this day.

\par
The standard approach in statistical mechanics is to identify thermodynamic state functions with specific average values  of a suitably chosen statistical ensemble that correctly reflects the physical conditions under which measurements are performed (perfect isolation, coupling to an  energy or matter reservoir, etc.). The most fundamental statistical ensemble is the microcanonical ensemble (MCE), describing the thermodynamics of isolated systems that are governed by energy conservation and which, at equilibrium,  cannot exchange heat or matter with their surroundings. The MCE is the foundation of other thermostatistical ensembles,  including the canonical ensemble (which permits permanent energy and/or heat exchange with the environment) and the grand-canonical ensemble (which allows both energy and matter exchange). These two subordinate  ensembles can be derived from the MCE by considering a subsystem of interest that is {\it weakly} coupled to the rest of the globally isolated microcanonical  system, which is then interpreted as an environment (heat bath or particle reservoir) for the particular subsystem.

\par %molecule pulling etc., Roldan Nature physic
Recent experimental advances make it possible to investigate thermodynamic properties of very small systems (single molecules, Brownian colloids or even individual atoms) that may be, in good approximation, decoupled from the environment or that can be in weak or strong contact with a much larger system. Such finite-system studies provide a valuable testbed for the notion and meaning of thermodynamic temperature in the context of various statistical ensembles. Particularly interesting from a theoretical and practical perspective are situations in which different ensemble descriptions are not guaranteed to be equivalent. Ensemble inequivalence is more norm than exception in finite-size systems but can also occur in macroscopic systems with long-range interactions or when the  density of states (DoS) is a nonmonotic function of energy. Equilibrium systems of the latter type are often classified as anomalous~\cite{Kubo} and,  if entropy is chosen naively, they can give rise to the paradoxical notion of a negative \lq absolute \rq{} temperature.

\par
In the remainder of this contribution,  we will survey the meaning of temperature in thermodynamics by summarizing and commenting on results from recent more detailed studies~\cite{HHD,Campisi}.

\section{Microcanonical thermodynamics and absolute temperature}

%This Law has also been used as the  starting point to construct uniquely the thermodynamic entropy of isolated systems in Ref. \cite{Campisi}.

The MCE describes the thermostatistics of a strictly isolated system through the density operator $\rho=\delta(E-H)/\omega$, where the normalization constant $\omega$ is the density of states (DoS). The MCE is the most fundamental ensemble as it only relies on the conservation of energy $E$, arising from the time-translation invariance of the underlying Hamiltonian $H$. External thermodynamic control parameters $Z$, such as available system volume, particle numbers, electric or magnetic fields,  enter as parameters through the Hamiltonian $H(Z)$ and the DoS~$\omega(E,Z)$. To connect the MCE to thermodynamics, J.~W.~Gibbs~\cite{Gibbs} studied two different candidates for the thermodynamic entropy of an isolated system.  The first is the {\it  volume entropy}, which in modern notation takes the form
\begin{equation}
\SG= \kB \ln {\Omega (E,Z)} \;. \label{volume}
\end{equation}
Here, $\kB$ denotes the Boltzmann constant  and the dimensionless {\it volume}-quantity $\Omega (E, Z)$ is the integrated DoS, classically obtained by integrating the non-negative  DoS $\omega\ge 0$ up to energy $E$,
\begin{equation}
\label{Omega}
  \Omega(E,Z)  = \int_0^E dE' \omega(E', Z),
\end{equation}
assuming zero ground-state energy for a physically stable system. Since $\Omega$ is non-decreasing function of $E$, the temperature $\TG$ obtained from~$\SG$ and Eq.~\eqref{e:T_def} is strictly non-negative.
\par
For classical Hamiltonian systems $H (\xi, Z)$ with phase-space variables $\xi$, the  integrated DoS $\Omega(E, Z)$ equals the properly normalized (via division by the symmetries of the degrees of freedom) and dimensionless (via division by the appropriate power of Planck's constant) integrated phase space volume up to the energy $E$. We may write this formally as
\begin{equation}
\label{Omega-b}
\Omega(E,Z)=  \Tr_\xi\,  \Theta[ E - H(\xi, Z)]
\end{equation}
where  $\Theta$  denotes the unit-step function and $\Tr$ the phase-space integral over distinguishable microstates~$\xi$.
For isolated quantum systems with discrete energy spectrum, $\xi$ comprises the complete set of quantum numbers, and we may interpret $\Omega$ in Eq.~\eqref{Omega-b} as a discrete level counting function, defined on the spectrum $\{E_n\}$ of the Hamiltonian. Intuitively, the discrete function $\Omega(E_n,Z)$ sums the eigenspace dimensions of the eigenvalues $E_j\le E_n$. In the quantum case,  one needs to postulate additional smoothing procedures before one can apply differential thermodynamic relations such as Eq.~\eqref{e:T_def} (see discussion in Sec.~\ref{s:open issues} below).
\par
Following Gibbs' seminal work,  Hertz \cite{Hertz} demonstrated the mechanical adiabatic invariance of the volume entropy $\SG$ for classical systems. The exact connection between $\SG$, its corresponding temperature $\TG$ and equipartition for classical finite size systems was emphasized in early works by Schl\"uter~\cite{Schluter} and Khinchin~\cite{Khinchin}. More recent discussions and applications of the Gibbs' volume entropy can be found in Refs.~\cite{HHD,Berdichevsky,Campisi2,PhysicaA,Casetti,Janke,DH,Baeten}.

\par
The second microcanonical entropy candidate studied by Gibbs~\cite{Gibbs} is the {\it surface entropy}
\begin{equation}
\SB = \kB \ln [\omega (E, Z)\epsilon].
\label{surface}
\end{equation}
The quantity $\epsilon$ denotes an arbitrary energy constant, needed to make the argument of the logarithm dimensionless.  That the definition of $\SB$ requires such an additional \textit{ad hoc} parameter is conceptually unappealing, but bears no relevance for thermodynamic  quantities that are related to derivatives of $\SB (E, Z)$ -- provided $\epsilon$ is assumed to be independent of $(E, Z)$. One can show however that the presence of $\epsilon$ can cause $\SB$ to violate Planck's formulation of the Second Law~\cite{HHD}. For discrete quantum systems with singular DoS $\omega$, Eq.~\eqref{surface} also requires additional interpolation and/or smoothing procedures (see Sec.~\ref{s:open issues}). The subscript \lq B\rq{} in Eq.~\eqref{surface} signals that this definition is also commonly referred to as Boltzmann entropy nowadays, which unfortunately does not seem to reflect properly the actual history. Although Boltzmann's tombstone famously carries the entropy formula $S = \kB \log W$,  it was, according to Arnold Sommerfeld \cite{Sommerfeld}, Max Planck \cite{Planck1,Planck2}  who first established this relation. As described in many textbooks, the entropy expression  $\SB$ in Eq. (\ref{surface}) can be heuristically obtained  by
identifying $\log = \ln$ and interpreting $W = \epsilon \omega(E, Z)$ as the number of microstates accessible to a physical system at energy $E$. This may explain the popularity of the term  \lq Boltzmann entropy\rq{}.
\par

It is well known that for macroscopic {\it normal} systems~\cite{Kubo} with a large number of microscopic degrees of freedom, most of the phase space volume is contained in a narrow shell just below the energy $E$. In such cases, the two entropy definitions become essentially indistinguishable and predict practically identical  thermodynamic equations of state.
There exists, however, a wide range of systems for which $\SB$ and $\SG$ are non-equivalent.

\subsection{Self-consistency checks}

The question as to whether $\SG$ or $\SB$ are viable candidates for the thermodynamic entropy of isolated systems, can be answered directly by testing either candidate against the Laws of Thermodynamics. The approximation-free analysis in Ref.~\cite{HHD} shows that
for a broad class of physical systems, which includes all standard classical Hamiltonian systems\footnote{These are confined systems with quadratic kinetic energy and finite ground-state energy.} of \textit{arbitrary} size, the Gibbs volume entropy $\SG$ satisfies the traditional formulations of the Zeroth, First and Second Law exactly. In contrast, the surface entropy $\SB$ is found to violate these laws in many situations~\cite{Gibbs,Campisi,HHD}. While referring the reader to Ref.~\cite{HHD} for technical details, we briefly summarize the most essential results.
\vspace{0.5cm}
\paragraph*{i.}
 That $\SG$, but not $\SB$, satisfies the Zeroth Law is a reflection of the fact~\cite{Schluter,Khinchin,Campisi2,PhysicaA,Baeten,HHD} that only $\SG$ satisfies the microcanonical equipartition theorem exactly. More precisely, denoting the microcanonical averages by $ \langle\,\cdot\, \rangle_E $, the Stokes theorem implies~\cite{Khinchin} implies that, for all  standard classical Hamiltonian systems with topologically simple phase space $\mathbb{R}^{d}$, the equipartition identity\footnote{For systems with complex phase space topology, Eq.~\eqref{e:equi} can be violated, see example in Sec.~\ref{s:example} below, where the phase space regions  corresponding clockwise and anti-clockwise motion become disconnected  for supercritical energy values. Such topologically peculiar systems do not thermalize in the traditional sense. However, for the most commonly considered standard classical Hamiltonian systems, Eq.~\eqref{e:equi} is strictly satisfied.} 
\begin{equation}
\kB \TG = \left(\frac{\partial \SG}{\partial E}\right)^{-1}  =
\left\langle    \xi_k \frac{\partial H}{\partial \xi_k} \right\rangle
\label{e:equi}
\end{equation}
holds for \textit{any} of the canonical coordinates $(\xi_1,\ldots, \xi_d)$. By contrast, this relation is \emph{in general} violated for the surface entropy, ruling out~$\SB$ as a consistent thermodynamic entropy. 

\vspace{0.5cm}
\paragraph*{ii.} Compliance of $S=S_G$ and $T=\TG$ with the First Law
\begin{equation}
\label{e:first_law}
dS=\frac{1}{T}dE+ \sum_i \frac{F_i}{T}dZ_i,
\qquad
F_i := T \Big (\frac{\partial S}{\partial Z_i}\Big )_E
\overset{!}{=} - \Big \langle \frac{\partial H}{\partial Z_i} \Big \rangle_E \;,
\end{equation}
follows directly from a simple integration by parts~\cite{HHD,DH}.  Note that the last equality in Eq.~\eqref{e:first_law} ensures that statistical averages agree with thermodynamic observables. One can easily verify that this consistency criterion is, in general, violated for the Boltzmann entropy $\SB$.

\vspace{0.5cm}
\paragraph*{iii.} Planck's Second Law of Thermodynamics for isolated systems can be, in essence, stated as follows:
Consider two isolated microcanonical systems that are initially separated and have entropies $S_1(E_1)$ and $S_2(E_2)$, respectively. Now couple the two systems weakly and let them equilibrate. Assuming energy conservation throughout the process, the joint equilibrated system is again microcanonical and has entropy $S_{1+2}(E_{1+2})=S_{1+2}(E_1+E_2)$.  Then, Planck's Second Law demands that the entropy of the final state is larger than the sum of the initial entropies,
\begin{equation}
\label{e:second_law}
S_{1+2}(E_{1+2})\ge S_1(E_1)+ S_2(E_2).
\end{equation}
Basic integral convolution properties imply that Eq.~\eqref{e:second_law} is always satisfied for $\SG$ (in most cases, even with strictly~\lq$>$\rq) but not necessarily by $\SB$ ~\cite{HHD}.
\par
In this context,  it is worthwhile to note that any subsequent attempt to decouple the two systems results in non-microcanonical distributions for the separated systems, since the exact individual energies are not known anymore due to the permanent energy exchange during the equilibration phase (i.e., thermal coupling is irreversible). This means that, without further manipulation or measurements (or the introduction of a Maxwell demon), the total entropy remains $S_{1+2}(E_1+E_2)$ after separation, a fact that has been missed by authors~\cite{Wang} who recently criticized the Gibbs entropy. Unsurprisingly, this basic error led to paradoxical conclusions~\cite{Wang}, such as an apparent violation of mathematically exact inequalities.

\vspace{0.3cm}
\par
For completeness, we mention that previous studies rarely focused on the Third Law, mainly because it is well known that many classical systems (including the ideal gas) do not obey the Third Law. Typically, verification of the Third Law requires a consistent quantum-mechanical  treatment\footnote{As a note of caution: One can find many partially conflicting versions of the Third Law in the literature, and some naive formulations are not applicable to isolated systems, or only apply to systems with non-degenerate ground-state or finite energy gap between ground-state and lowest excited energy levels.}. Evidently, the Gibbs entropy satisfies $\SG(E_0)= \kB \ln g_0$ with $g_0$ denoting the degeneracy of the ground state energy $E_0$ and hence fulfills the most basic version of the Third Law.

\subsection{Positive and negative absolute temperatures: an example}
\label{s:example}
\begin{figure*}[t]
\centerline{
\includegraphics[height= 4.5cm]{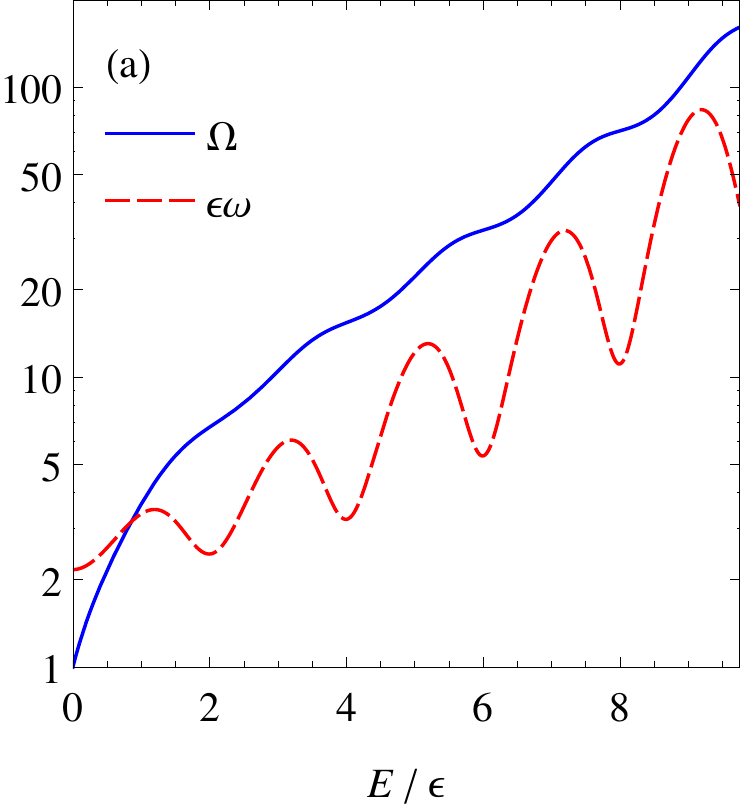}
\hspace{0.5cm}
\includegraphics[height= 4.5cm]{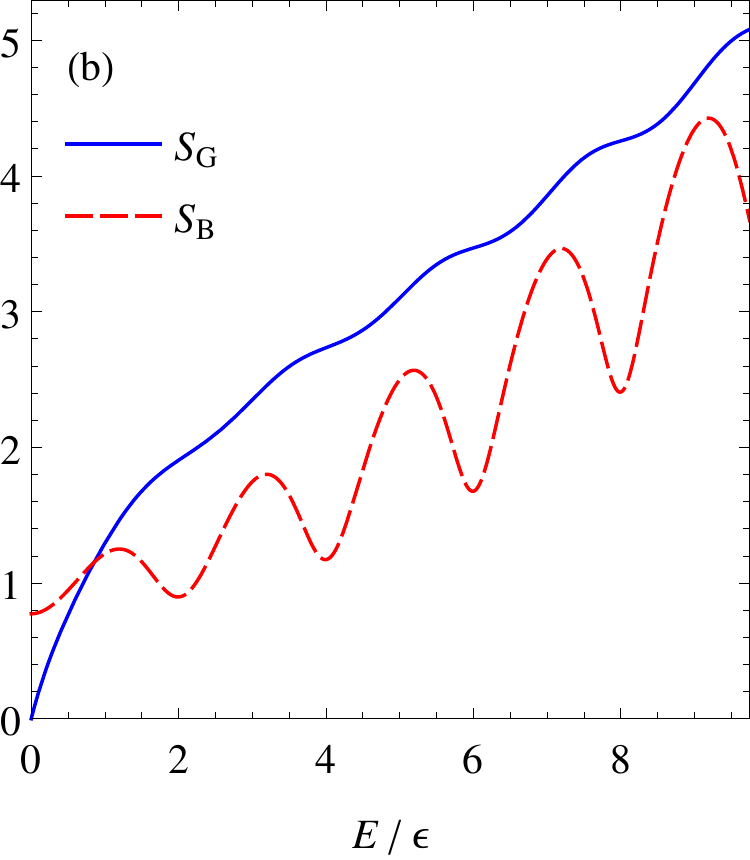}
\hspace{0.5cm}
\includegraphics[height= 4.5cm]{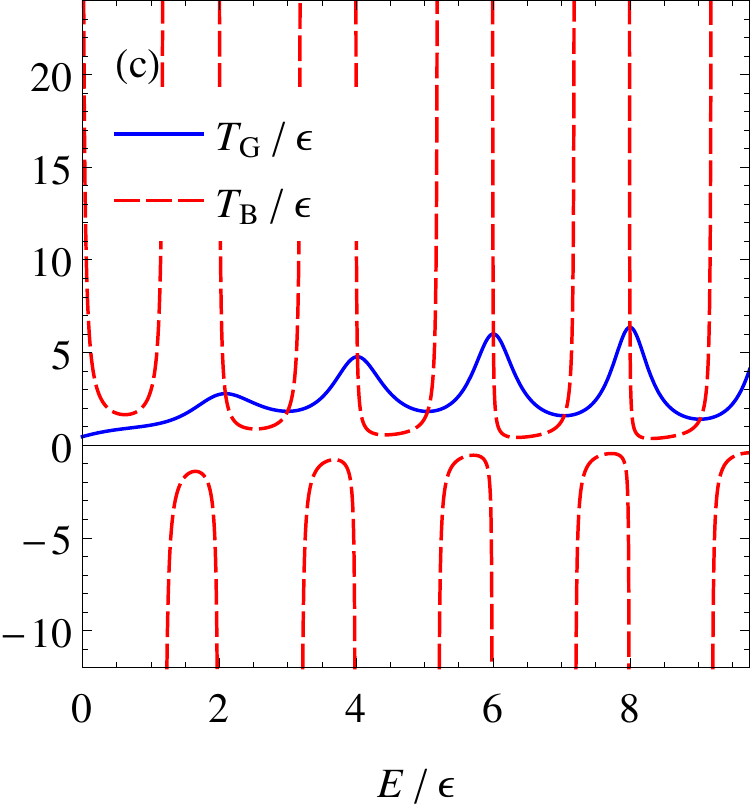}
}
\caption{
\label{fig:wiggly_dos}
Non-uniqueness of microcanonical temperatures for a system with nonmonotonic density of states (DoS); figure adapted from Ref.~\cite{HHD}. (a):~DoS $\omega$ (red, dashed) and integrated DoS $\Omega$ (blue) for the example in Eq. (23) of Ref.~\cite{HHD}. (b):~Gibbs volume entropy $\SG$ (blue) and Boltzmann surface entropy $\SB$ (red, dashed) in units $\kB=1$. (c):~Gibbs temperature~$\TG$ (blue) and Boltzmann temperature~$\TB$ (red, dashed). The example illustrates that, in general, neither the Gibbs nor the Boltzmann temperature uniquely characterize the thermal state of an isolated system because the same temperature value can correspond to different energy values.
}
\end{figure*}

The primary thermodynamic state variables of an isolated system with Hamiltonian $H(Z)=E$ are energy $E$ and control parameters $Z$.  By contrast, the temperature $T$ is a secondary derived quantity determined by Eq.~\eqref{e:T_def}. For the Gibbs volume entropy, one finds explicitly
\begin{equation}
\label{T_G}
\kB \TG = \frac{\Omega(E, Z)}{\omega(E, Z)},
\qquad
\omega(E, Z)  =\frac{\partial \Omega}{\partial E}.
\end{equation}
Since both the integrated DoS  $\Omega\ge 0$ and the DoS $\omega\ge 0$ are non-negative,
the Gibbs temperature is strictly non-negative. For comparison, the Boltzmann temperature is given by
\begin{equation}
\label{T_B}
\kB \TB   = \frac{\omega(E, Z)}{\nu(E, Z)} ,
\qquad
\nu(E, Z)  =\frac{\partial \omega}{\partial E}.
\end{equation}
The Boltzmann temperature $\TB$ is negative whenever the DoS $\omega$ is a locally decreasing with $E$ (see example in Fig.~\ref{fig:wiggly_dos}).  This happens for Ising-type spin or laser systems in the population inverted system,  as well is in Hamiltonian systems exhibiting singular points in their DoS that separate regions of $\nu(E, Z)>0$ with regions with negative-valued $\nu(E, Z)$.

\par
An instructive ergodic 1D example~\cite{Baeten} is the classical pendulum (mass $m$, length $L$, gravitational acceleration $g$) with Hamiltonian
\begin{equation}
H(\phi, p_{\phi}) = \frac{p_{\phi}^2}{2m} - mgL\cos \phi.
\end{equation}
The energy range is bounded from below but unbounded from above,  $0<E<\infty$. The DoS $\omega$  can be given analytically in terms of  complete elliptic integrals of the first kind (see in Ref.~\cite{Baeten} and Fig. 1 therein). The DoS increases in the oscillatory regime $0<E<  mgL$, exhibits a singularity at $E= mgL$, where the orbit period diverges, and decreases in the continuous-rotation regime, $mgL < E < \infty$. Accordingly, \mbox{$\nu(E > mgL)<0$} decays monotonically towards zero as $E\rightarrow \infty$, implying a negative Boltzmann temperature for $E > mgL$.  By contrast, the Gibbs temperature is positive for all $E>0$. In particular, for $E\gg mgL$, any further increase of energy is, in essence, purely kinetic and the system approaches an ideal one-particle gas on a circle, which should asymptotically satisfy $E=+\frac{1}{2}\kB T$, unless one is willing to give up this standard caloric equation of state. It easy to check that this relation holds only for $T=\TG$.

%================================================
%================================================

\subsection{Additional remarks}

\textbf{Heat does not always flow from hot to cold. }
The thermodynamic state of an isolated system is completely determined by the primary state variables~$(E,Z)$.  Since temperature is not a primary state variable, it cannot, in general, uniquely characterize the thermodynamic state of a microcanonical system (see Fig.~\ref{fig:wiggly_dos}).
This means there exist situations where neither the Gibbs temperature $\TG$ nor the Boltzmann temperature $\TB$ can predict the energy flow between weakly coupled systems that had different temperatures  before contact~\cite{HHD}. For instance, for a system with  oscillatory DoS,  two or more significantly different energy values can have the same temperature, regardless of which entropy definition one adopts (see Fig.~\ref{fig:wiggly_dos}c). Therefore,  the naive formulation of the Second Law \lq heat always flows from hot to cold\rq{} does not hold in general. Likewise invalid are versions of the Zeroth Law that claim that isolated systems with equal temperatures should not produce a net heat flow between them when brought into thermal contact. This can again be readily seen by considering, for example,  the coupling of an ideal gas to a system with oscillatory DoS as in Fig.~\ref{fig:wiggly_dos}.

\textbf{Thermodynamics applies to equilibrium systems of any size.}
It takes little effort to verify that Eqs.~\eqref{e:equi}-\eqref{e:second_law} hold exactly for standard  classical Hamiltonian systems with an arbitrary number of degrees of freedom~$N$. Similarly, the canonical ensemble discussed below can be applied to (sub-)systems of any size. These mathematical facts are widely appreciated by many colleagues~\cite{Janke,Casetti,Baeten,Campisi,Berdichevsky} -- in particular those interested in understanding DNA folding~\cite{Alemany}, microscopic information storage and erasure~\cite{Lutz_Nature}  and fluctuation phenomena~\cite{Roldan} --  and yet remain ignored by others~\cite{Wang,SwendsenWang}. When judged objectively, there is no doubt that the application of thermodynamic concepts to finite systems has  considerably advanced our understanding of biophysical, colloidal and quantum processes\footnote{Most of the experimental applications involve the canonical ensemble, as DNA molecules~\cite{Alemany} or colloids ~\cite{Lutz_Nature,Roldan} are typically held in a liquid bath that acts as a canonical thermostat. However, if one accepts the applicability of canonical thermostatistics to finite systems, then there exist no mathematical or logical or physical reasons that forbid the application of microcanonical statistics to finite systems, because Eqs.~\eqref{e:equi}-\eqref{e:second_law}  hold for systems of any size. Of course, the thermostatistical characterization of small systems should not just be limited to the mean values appearing in Eqs.~\eqref{e:equi} and \eqref{e:first_law} but should also include a careful fluctuation analysis of the underlying stochastic observables.}. Compared with infinite-systems thermodynamics, a practical difference is given by the fact that fluctuations generally play a (much) more important role in small systems. The presence of fluctuations, of course, does not mean that it is forbidden to characterize single DNA molecules thermodynamically; on the contrary, such fluctuations typically contain  important additional thermodynamic and energetic information that is usually lost in the infinite-system limit. Therefore, it would seem wiser to focus on understanding better the fluctuations of thermostatistical variables in finite systems, such as those of virial quantities on the rhs. of Eq.~\eqref{e:first_law}, instead of discarding finite-system thermodynamics on purely habitual grounds~\cite{SwendsenWang}.  Dogmatic insistence on the thermodynamic limit $N\to\infty$ is about as useful as insisting on the Newtonian limit, corresponding to speed of light $c\to \infty$, in relativity.  In both cases, things may become simpler, but one is missing out on relevant physics.

\textbf{Clausius relation \& Carnot efficiency.}
Campisi~\cite{Campisi} showed recently that the inverse Gibbs temperature $T_{G}^{-1}$  appears  naturally as the integrating factor in the Clausius relation for virtually all practically relevant physical systems. This corroborates the fact that the Gibbs temperature $\TG$ should be identified with the {\it absolute}  thermodynamic temperature $T$, unless one is willing to abandon the Clausius relation. Moreover, the non-negativity of the Gibbs temperature directly implies that
Carnot efficiencies cannot exceed 1.

\pagebreak
\textbf{Thermodynamic potentials can be \lq nonlocal\rq.}
It is sometimes argued that the Gibbs entropy cannot be the \lq correct\rq{} thermodynamic entropy as it is based on the integrated 
phase-space volume $\Omega$, which is a \lq nonlocal\rq~quantity that arises from a summation over states in an extended energy range.  This argument might appear superficially appealing but it is
ill-founded for (at least) two reasons: First, the microcanonical averages appearing in Eqs.~\eqref{e:equi} and~\eqref{e:first_law} are computed purely locally on the energy surface in phase space. Yet, the Stokes theorem implies that they can be related to the enclosed phase-space volume~\cite{Khinchin} and, hence, entropy should be a nonlocal volume-related quantity.  Second, if we insisted on purely local potentials everywhere in physics then we would have to stop using force potentials in mechanics, which are in essence \lq{}non-local\rq{} integrals over local forces experienced by particles. At this point, however, it is helpful to recall why such non-local potentials are introduced in mechanics in the first place: They allow us to define an important conserved quantity,  \emph{energy}.  Just as the energy is invariant under infinitesimal time translations, the integrated phase-space volume $\Omega$ is invariant under  infinitesimal adiabatic parameter translations~\cite{Hertz}. Hence, it should not be surprising but rather  be expected that thermodynamic potentials may be non-local in energy space.

\textbf{Ising models are bad benchmarks.}
When using specific theoretical models to illustrate alleged pros and cons of certain entropy
definitions, then it is advisable to verify first that these models respect superordinate  experimentally established knowledge. Specifically, while the observed stability of matter implies the existence of lower energy bounds on Hamiltonians, there exists no evidence to date for strict upper energy bounds. This means that $E\to -E$ is not a fundamental symmetry of physics and, hence, one should not impose such energy-reflection symmetry on thermodynamic quantities. For the same reason, it is not advisable to base arguments exclusively on Ising-type models, which are \textit{ad hoc} truncations of more fundamental Hamiltonians that are \emph{not} bounded from above, if one wants to evaluate the conceptual validity of a certain thermostatistical framework.

\textbf{Ensemble (in)equivalence.}
 Although $\SG, \SB$ and other entropy candidates \cite{HHD} often yield practically indistinguishable predictions for the thermodynamic properties of large {\it normal} systems \cite{Kubo}, such as quasi-ideal gases with macroscopic particle numbers, they can produce substantially different predictions for finite mesoscopic systems, for ad-hoc truncated Hamiltonians with upper energy bounds or even for macroscopic gravitational systems~\cite{Gross}. This implies (see discussion in Sec.~\ref{sec:canonical} below) that, in general, microcanonical and canonical descriptions are not equivalent, which is not surprising as they refer to different physical conditions (complete isolation \textit{vs.} coupling to an infinite bath).

%%%%%%%%%%%%%%%%%%%%%%%%%%%%%
\section{Temperature in the canonical ensemble }
\label{sec:canonical}
%%%%%%%%%%%%%%%%%%%%%%%%%%%%%

The canonical Boltzmann factor $e^{-\beta H}$, where $T=(\kB\beta)^{-1}$ is commonly identified with the bath temperature, has become one of the most frequently employed statistical tools in physics. It is therefore conceptually important and practically useful to understand potential validity limits, which arise from  assumptions and approximations made during the derivation from the underlying MCE.

\subsection{Boltzmann factor and temperature}

We briefly summarize the key assumptions underlying the derivation of the Boltzmann factor by considering a system of interest $\mathcal{S}$ which is coupled to another system $\mathcal{B}$ that acts as a heat bath. The starting point of the derivation is the microcanonical density operator ${\rhoT(\xi|\ET, Z)= \delta[\ET - \HT(\xi, Z)]/ \omegaT(\ET, Z)}$ of the total system $\mathcal{T}=\mathcal{S} + \mathcal{B}$.  The first key assumption en route to the Boltzmann factor is \emph{weak coupling}, which means that one neglects  system-bath interaction contributions to the total energy and Hamiltonian, by writing~$\ET= \ES + \EB$  and $\HT = \HS + \HB$. Because the total energy $\ET$ is fixed, the probability weight $P(\ES| \ET,Z)$ to find the fluctuating energy value $\ES$ of the subsystem $\mathcal{S}$ is simply given by~\cite{Kubo,Callen, HHD}
\begin{equation}
\label{weight-0}
P(\ES| \ET,Z) = \frac {\gS(\ES)\, \omegaB(\ET - \ES)}{\omegaT(\ET)}
\end{equation}
Here, $\gS(\ES)$ denotes the degeneracy of the subsystem energy value $\ES$, while $\omegaB(\EB)$ is the DoS of the bath at energy $\EB = \ET - \ES$, and $\omegaT(\ET)$ the total DoS at the total energy $\ET$.  Equation~\eqref{weight-0} can be equivalently rewritten as
\begin{equation}
\label{weight}
P(\ES| \ET,Z)      =  \frac {\gS(\ES)}{\epsilon\,\omegaT(\ET)} \exp \left[\frac{\SBB(\ET-\ES)}{\kB}\right]
\end{equation}
where $\SBB(\EB)=\SBB(\ET-\ES)$ denotes the Boltzmann entropy of the \emph{bath}. As next step in the standard derivation \cite{Kubo,Callen}, one expands the Boltzmann entropy in the exponent around some conveniently chosen value $\EBE$, typically taken to be  the expectation value $\EB$ of the bath energy\footnote{Replacing the expansion point $\EBE$ by the mode is not recommendable, as this procedure becomes ambiguous or even ill-defined  when the DoS $\omegaB(\EB)$ of the bath is oscillating or  monotonically increasing.}, keeping terms up to linear order:
\begin{equation}
\label{e:expansion}
\SBB(\ET -\ES) =
 \SBB(\EBE) + \frac{1}{\TBB(\EBE)}(\ET - \ES - \EBE) + \ldots,
\end{equation}
where  $\TBB = (\partial \SBB /\partial \EB)^{-1}$ is Boltzmann temperature of the bath.
Note that Eq.~\eqref{e:expansion} is essentially an expansion in the energy fluctuations of the bath $\delta E^B=\EB-\EBE = (\ET - \ES) - \EBE$. Inserting the expansion~\eqref{e:expansion} into Eq.~\eqref{weight} gives\footnote{
The approximation \eqref{approx_weight} neglects all higher-order contributions in the Taylor expansion of $\SBB$. For example, the coefficient in front of the  quadratic term is proportional to
$\partial^{2} \SBB/\partial^2  {\EB} = -1/({\TB}^2 \CBB)$, where $\CBB= \partial \EB/\partial \TB$ is the canonical specific heat of the bath. For this term to vanish individually,  $\CBB$ has to be sufficiently large. Roughly speaking, one can expect that the second-order as well as higher-order expansion terms become negligible if the Boltzmann temperature $\TBB$ changes only slowly when the bath energy is varied. This typically requires a large bath.
}
\begin{equation}
\label{approx_weight}
P(\ES| \ET,Z)  = \frac {\gS(\ES)}{\epsilon\,\omegaT(\ET)}\exp \left[\frac{\SBB(\EBE)}{\kB} +
\frac{(\ET - \EBE) - \ES }{\kB \TBB(\EBE)} +  \ldots \right].
\end{equation}
Assuming~\emph{all} higher-order terms can be neglected,  one obtains the standard result
\begin{equation}
\label{Boltzmann}
P(\ES| \ET,Z)   = \frac {\gS(\ES)}{\ZC}\exp \left[-\frac{\ES}{\kB \TBB(\EBE)} \right] .	
\end{equation}
where all remaining $\ES$-independent terms have been absorbed into the normalizing constant $\ZC$, which is the canonical partition function.\par
Thus, the temperature entering the celebrated Boltzmann factor $\exp(-\beta \ES)$ is the {\it Boltzmann temperature}  $\TBB$ of the {\it bath}.  One may therefore be tempted to assume that
$\TBB$ can be identified with the thermodynamic temperature of the total system. However, this is logically incorrect because $T^\mathcal{B}_B(\bar{E}^\mathcal{B})$ is in general not equal~\cite{HHD} to the total system Boltzmann temperature $T^\mathcal{T}_B({E}^\mathcal{T})$ or Gibbs temperature $T^\mathcal{T}_G({E}^\mathcal{T})$, with the latter being the actual thermodynamic temperature. Of course, when the bath is macroscopically large and normal (e.g., ideal gas-like) from a thermostatistical viewpoint, then the effective temperature $\TBB$ practically coincides with the Gibbs temperatures of the bath $\mathcal{B}$, the system~$\mathcal{S}$, and the total system~$\mathcal{T}$,  and we have  $\TBB =\TGB =\TGS =\TGT$ in this limit. In contrast, when considering finite thermostats (i.e., a bath  with a finite number of microstates), then the exponential Boltzmann has to be replaced by a generalized Boltzmann factor, which may assume the form  of a Tsallis-Renyi escort distribution \cite{CampisiPRE2009}.

\subsection{Beyond weak coupling}

Having surveyed the notion of temperature in the canonical ensemble, we still address another subtlety that concerns the heat capacities of nano-systems. When studying the thermodynamic properties of nano-scale devices, one naturally encounters the question whether the weak coupling approximation remains justified. Indeed, for a typical nano-system in contact with a heat bath, the coupling energy is usually of the order of the average system energy. Therefore, coupling terms in the Hamiltonian can no longer be neglected. This issue was investigated in detail in Refs.~\cite{negC1a,negC1b,negC2}, which focused on the question how the canonical heat capacity is affected when nano-subsystems are \emph{strongly} coupled to a large normal bath.  These studies showed that the reduced canonical weight (or reduced density operator in quantum mechanics) of a strongly coupled small system is no longer of the Boltzmann form, when expressed in terms of the bare sub-system energy  $\ES$ or corresponding Hamilton operator $\HS$. Instead, the canonical weight now features a renormalized subsystem Hamiltonian that  depends explicitly on both effective bath temperature $\TB$ and coupling strength. Moreover, as a main consequence, the canonical specific heat of the subsystem is not guaranteed to be positive  and can, in fact, attain {\it negative} values even for $\TB>0$~\cite{negC1a,negC1b}. The thermodynamic entropy\footnote{For non-weakly coupled systems, correlations between bath and system cannot be neglected and, therefore, the thermodynamic entropy of  such subsystems is no longer given by  the classical Gibbs-Shannon entropy or the quantum-mechanical von-Neumann entropy.} of such a strongly coupled quantum system, obtained from its canonical partition function via the free energy, assumes a form that is close (but not exactly equal) to the quantum conditional entropy, and can become negative for $\TB>0$~\cite{negC2}.  This does not affect, however, the validity of the Third Law as stated above, which holds true for $\TB\to 0$ even when a small quantum subsystem is strongly coupled to a heat bath; see Fig.~3 in  Ref.~\cite{negC2} for an example.

\subsection{Thermodynamic \textit{vs.} information entropy}
We conclude our discussion of the canonical ensemble with brief remarks on thermodynamic and information-theoretic entropies. The exponential Boltzmann distribution~\eqref{Boltzmann}
is directly linked to the  entropy $\SC = - \kB {\Tr} (\rho \ln \rho)$, as already noted by Gibbs~\cite{Gibbs} who discussed $\SC$ exclusively in the context of the canonical ensemble. Nowadays, $\SC$ is commonly referred to as the {\it canonical} Gibbs-Shannon entropy in classical statistical mechanics and as the von Neumann entropy in quantum statistics.  It is well-known that the canonical distribution~\eqref{Boltzmann} can be obtained by maximizing $\SC$ under the assumption that the mean energy $\bar{E}^\mathcal{S}$ is given. However, such a purely formal \lq derivation\rq{}  conceals the underlying physical assumptions that determine the range of validity of the Boltzmann distribution~\eqref{Boltzmann}. Also, entropy maximization arguments often leave the impression that there is a direct 1-to-1 correspondence between thermodynamics and information theory, which is somewhat misleading  for a number of reasons: First, there exist many different information measures~\cite{Renyi} and the Shannon entropy is just one of them - although an admittedly very nice one. Second, the Shannon entropy can be used to quantify the information content of arbitrary probability measures that, in most cases, have no relation to the thermodynamic equilibrium distributions. Third, the most fundamental equilibrium ensemble, the MCE, has a thermodynamic  entropy that does not belong to the class of Shannon entropies.   Therefore, some reservation seems in order when attempts are made to identify information-theoretic measures generically with thermodynamic entropies and \textit{vice versa}. A similar note of caution applies when one tries to relate information-theoretic inequalities to thermodynamic inequalities that arise in the context of the Second Law or from thermodynamic stability considerations~\cite{Callen}. Potential analogies between thermodynamics and information theory are interesting and deserve to be explored in  great detail, but they should not necessarily be raised to the level of postulates, when they have been shown to be incomplete and may obscure physical insight.

\section{Open questions}
\label{s:open issues}
The above discussion implicitly assumed that all derivatives exist and are well behaved. This is typically the case for classical Hamiltonian systems with the exception of critical points~\cite{PhysicaA}, as also encountered in the pendulum example above~\cite{Baeten}.  For quantum systems, the problem is generally more subtle since quantum-mechanical  energy spectra can be partially or completely discrete, and  are typically very sensitive to small perturbations that can break symmetry-related degeneracies.  Similar effects occur in classical approximations to quantum systems, as for example the classical Ising model.  Whenever one faces a completely or partially discrete spectrum $
\{E_i\}$, the corresponding  DoS $\omega(E,Z)$ becomes formally singular and essentially reduces to a collection of $\delta$-function at those discrete energy values, $\omega(E)=\sum_i g_i\delta(E-E_i)$. In this case, the construction of a differentiable DoS  requires some sort of smoothing procedure. This issue is closely related to the so-called \lq\lq Weyl problem\rq\rq{} of finding asymptotic  approximations for the eigenvalue distributions of Hermitian operators in finite domains,   by applying some suitable averaging procedure to obtain a continuous DoS~\cite{Baltes}. For canonical systems, one typically uses such a smoothed DoS of the underlying energy spectrum at high ambient temperatures.
\par
When the spectrum exhibits a discrete range, then one can define the integrated DoS $\Omega(E, Z)$, which enters the microcanonical Gibbs entropy $\SG$,  in at least two different ways: The most commonly used method simply integrates the discrete DoS $\omega(E,Z)$ over $E$, which results in a step function $\tilde{\Omega}$ that gives rise to singular thermodynamic derivatives. This approach seems unsatisfactory mathematically, for one simply integrates over the 'forbidden' part\footnote{The union of intervals $(E_i,E_{i+1})$.} of the spectrum while completely ignoring structural information encoded in the amplitude values $g_i$ of $\omega$ in the interpolation regions $(E_i, E_{i+1})$. A potentially better method~\cite{HHD} is based on analytic interpolation of the discrete level counting function~\mbox{$\Omega(E_n)=\sum_{E_j\le E_n}\dim \mathcal{H}_j$}, where $\mathcal{H}_j$ is the eigenspace of $E_j$. Although the most natural interpolations appear obvious when $\Omega(E_n)$ can be written as $\Omega(E_n)=f(n)$ for some known function $f$~(see examples in Ref.~\cite{HHD}), there remain open questions as to how to treat rigorously cases where no such closed-form representation is known.\footnote{Another practical, but not quite as elegant approach is to replace derivatives with finite differences, which in essence corresponds to linear and higher-order polynomial interpolations.}

\section{Conclusions}

Gibbsian thermodynamics~\cite{Gibbs,Hertz} works consistently for finite and infinite systems, because the underlying mathematical and statistical foundations, most importantly Liouville's theorem,  merely rely on generic conservation laws that arise from the Hamiltonian structure of the microscopic dynamics. Working with infinite systems is generally easier as this limit often (but not always) forgives a certain laxness in defining entropy and thermostatistical observables, since different definitions may show the same asymptotic behavior when the particle number $N$ is let to $\infty$.  As mentioned before, this is quite analogous to the fact that the Newtonian limit $c\to\infty$ is generally easier to handle than a full relativistic treatment at finite speed of light~$c$, which of course does not mean that Newtonian dynamics is more correct than relativity. Just as relativity compels us to think more carefully about how to formulate fundamental physical laws, the thermostatistical analysis of finite systems forces us to pay more rigorous attention to mathematical and physical consistency criteria~\cite{HHD} in thermodynamics. This profound insight can be attributed to Gibbs, who wrote on page 179 of his fundamental treatise~\cite{Gibbs}: \lq\lq It would seem that in general averages are the most important, and that they lend themselves better to analytical transformations. This consideration would give preference to the system of variables in which $\log V$ [$=\SG$ in our notation] is the analogue of entropy. Moreover, if we make $\phi$  [$=\SB$ in our notation] the analogue of entropy, we are embarrassed by the necessity of making numerous exceptions for systems of one or two degrees of freedoms.\rq\rq{} Gibbs was, of course, well aware that statistical fluctuations become important in finite systems and that, therefore, the exact thermodynamic mean value relations~\eqref{e:equi} and~\eqref{e:first_law} have to be complemented by detailed fluctuation analysis, as nowadays the norm in DNA and colloid experiments~\cite{Alemany,Lutz_Nature,Roldan}. 
\par
In this contribution, we have surveyed the notion of thermodynamic temperature in the microcanonical and the canonical ensemble. For isolated microcanonical systems, the Gibbs volume entropy fulfills exactly the standard laws of thermodynamics as well as equipartion for a wide range of systems, including all classical standard Hamiltonian systems regardless of their size. For finite systems, fluctuation analysis provides important physical information beyond the mean values that define standard thermodynamic state-variables.  The microcanonical Gibbs formalism agrees with the Clausius relation~\cite{Campisi}, implies strictly non-negative temperatures and, hence, ensures Carnot efficiencies $\le 1$. By contrast, the Boltzmann entropy, which can yield  \lq negative absolute temperatures\rq,  is not a consistent thermodynamic entropy if one adopts the standard Laws of Thermodynamics, as summarized in Eqs.~\eqref{e:equi}--\eqref{e:second_law}.
It is therefore not obvious to us why one should favor an entropy that can violate Planck's law~\eqref{e:second_law} over one that fulfills it rigorously.
\par
Notwithstanding, the Boltzmann temperature plays an important role as an effective bath temperature in the canonical ensemble, describing a subsystem that is in weak contact with a quasi-infinite environment. If the bath behaves normally (e.g., ideal gas-like), then the Boltzmann temperature practically coincides with the Gibbs temperature. Subtle differences arise, however, for systems that are non-weakly coupled to  an environment, as typically the case for nano-scale devices.  In the presence of strong coupling, the specific heat of the device can become negative \cite{negC1a,negC1b} even though the total system consisting of device and bath is thermodynamically stable. This feature is in stark contrast to the weak-coupling case, where the canonical specific heat of the subsystem is  strictly positive.

\par
Last and least, some authors~\cite{Wang,SwendsenWang,Wang2} have recently criticized the microcanonical Gibbs formalism~\cite{Gibbs,Planck2} by limiting their discussion to infinite systems and advocating modified versions of the thermodynamic laws, tailored to favor their own preferred entropy definitions.
If one accepts such reasoning, then one must also be willing to replace the exact Eqs.~\eqref{e:equi}--\eqref{e:second_law} with inexact approximations -- which seems a steep price to pay. The exactness of Eqs.~\eqref{e:equi}--\eqref{e:second_law} is not a consequence of specific postulates but follows from basic integral and differential calculus (the \lq proofs\rq{} are trivial and take only a few lines~\cite{HHD}). Hence, even if one dislikes the Gibbs formalism as developed in Refs.~\cite{Gibbs,Hertz,Planck2}, one should at least acknowledge the correctness of the mathematically rigorous results~\eqref{e:equi}--\eqref{e:second_law}. Moreover, instead of focusing on the discussion of abstract postulates~\cite{SwendsenWang},  it may also be useful to remind ourselves that the purpose of any thermodynamic theory should be the prediction of physically measurable quantities, such as pressure, magnetization, etc. which correspond to operationally well-defined statistical averages. We would therefore encourage readers who care about the practical applicability of theoretical concepts to perform the following simple numerical experiment: 
\begin{enumerate}
\item 
Place $N_1$ heavy particles (mass $m_1$) and $N_2$ light particles\footnote{Two particle types are required to make the dynamics sufficiently ergodic, if $N=N_1+N_2>1$.} (mass $m_2<m_1$) randomly on a finite 1D  interval $[0,L]$, and
 assign to each particle some initial velocity, corresponding to some total energy $E=(1/2)(\sum_{i=1}^{N_1} m_1 v_i^2 +\sum_{j=1}^{N_2} m_2 v_j^2) $.
\item
Evolve the system by assuming elastic momentum-conserving point-particle collisions  and total reflection at the interval boundaries.
\item
Measure the kinetic temperatures $\kB \bar{T}_s=\langle m_s v_i^2\rangle$ for each species $s=1,2$ and the kinetic pressure $\bar{p}$, conventionally defined  as the mean momentum transfer  to the interval boundaries per unit time, by taking standard time averages. 
\item
Compare $\bar{T}_s$ and $\bar{p}$ with predictions from the various entropy definitions for any combination $(N_1,N_2)$, starting with $(1,0), (0,1), (1,1),\ldots$.
\end{enumerate}
If one still intends to discard the Gibbs entropy afterwards, then one will have to explain why it is common sense to replace an entropy definition that produces correct predictions for all combinations $(N_1,N_2)$ by another one that does not.

\textbf{Acknowledgements}
The authors thank Michele Campisi, Gert-Ludwig Ingold and Peter Talkner for numerous insightful comments and discussions on these topics over the last ten years. This work was supported by the DFG Cluster of Excellence \lq Nanosystems Initiative Munich\rq{} (P.H.) and DFG Cluster of Excellence \lq{}Origin and Structure of the Universe\rq{} (S.H.) and an MIT Solomon Buchsbaum Fund Award (J.D.).

%%%%%%%%%%%%%%%%%%%%%%%%%%%%%%%%%%%%%%%%%%%%%%%%%%%%%%%%%%%%%%%%%%%%%%%
\end{document}